\documentclass[12pt]{article}
\usepackage{amsmath,amssymb,bm,epsfig}

\renewcommand{\thefootnote}{\fnsymbol{footnote}}

\begin{document}

\title{
\begin{flushright}
\begin{minipage}{0.2\linewidth}
\normalsize
WUF-HEP-15-13 \\
EPHOU-15-011 \\*[50pt]
\end{minipage}
\end{flushright}
{\Large \bf 
{Gauge coupling unification in 
$SO(32)$ heterotic string theory with 
magnetic fluxes
\\*[20pt]}}}

\author{Hiroyuki~Abe$^{1,}$\footnote{
E-mail address: abe@waseda.jp},\ \ 
Tatsuo~Kobayashi$^{2}$\footnote{
E-mail address:  kobayashi@particle.sci.hokudai.ac.jp}, \ \ 
Hajime~Otsuka$^{1,}$\footnote{
E-mail address: hajime.13.gologo@akane.waseda.jp
},  \\
Yasufumi~Takano$^{2}$\footnote{
E-mail address: takano@particle.sci.hokudai.ac.jp}, \ and \
Takuya~H.~Tatsuishi$^{2}$\footnote{
E-mail address: t-h-tatsuishi@particle.sci.hokudai.ac.jp}
\\*[20pt]
$^1${\it \normalsize 
Department of Physics, Waseda University, 
Tokyo 169-8555, Japan} \\
$^2${\it \normalsize 
Department of Physics, Hokkaido University, Sapporo 060-0810, Japan} \\*[50pt]}

\date{
\centerline{\small \bf Abstract}
\begin{minipage}{0.9\linewidth}
\medskip 
\medskip 
\small
We study $SO(32)$ heterotic string theory on torus with magnetic fluxes.
Non-vanishing fluxes can lead to non-universal gauge kinetic functions 
for $SU(3) \times SU(2) \times U(1)_Y$ which is the important features 
of $SO(32)$ heterotic string theory in contrast to the $E_8\times E_8$ theory. 
It is found that the experimental values of gauge couplings are realized with 
${\cal O}(1)$ values of moduli fields based on the realistic models with the 
$SU(3) \times SU(2) \times U(1)_Y$ gauge symmetry and three chiral 
generations of quarks and leptons without chiral exotics.
\end{minipage}
}

\begin{titlepage}
\maketitle
\thispagestyle{empty}
\clearpage
\tableofcontents
\thispagestyle{empty}
\end{titlepage}

\renewcommand{\thefootnote}{\arabic{footnote}}
\setcounter{footnote}{0}
\vspace{35pt}

\section{Introduction}

Gauge coupling unification is a familiar tool to 
search for the underling theory of the standard 
model such as the Grand unified theory (GUT) or 
string theory.
For example, in the light of the observed values of gauge 
couplings, the three gauge couplings are unified 
with the GUT normalization of $U(1)_Y$ at the so-called GUT scale, 
$2\times 10^{16}$ GeV in 
the low-energy or multi-TeV scale 
 minimal supersymmetric standard model (MSSM).
 On the other hand, the three gauge couplings are not unified with the 
 GUT normalization of $U(1)_Y$  in the standard model (SM).

Superstring theory also has a certain prediction.
In particular, in 4D low-energy effective field theory derived from 
heterotic string theory, the gauge couplings at tree level are unified up to Ka\v{c}-Moody levels $\kappa_a$
at the string scale \cite{Ginsparg:1987ee}, 
which is of ${\cal O}(10^{17})$ GeV \cite{Kaplunovsky:1987rp}.
This prediction is very strong.
In order to explain the experimental values, 
we may need some corrections, e.g. stringy threshold corrections \cite{Dixon:1990pc,Antoniadis:1991fh,Derendinger:1991hq}.
(See for numerical studies e.g. Refs.~\cite{Ibanez:1991zv,Kawabe:1994mj}.)
Then, we may need ${\cal O}(10)$ of moduli values.\footnote{See for a recent study \cite{Bailin:2014nna}, where 
it was pointed out that ${\cal O}(1)$ of moduli values can be sufficient.}
When some moduli values are of ${\cal O}(10)$ or larger, the string coupling may become strong and 
peturbative description may not be valid.

On the other hand, in D-brane models, gauge couplings seem to be independent of each other 
for gauge sectors, which originate from different sets of D-branes.
Thus, one may be able to fit parameters such as moduli values in order to explain 
the experimental values of three gauge couplings, 
although there may appear some relations and/or constraints \cite{Blumenhagen:2003jy,Hamada:2014eia} 
in a certain type of models.

Here, we study another possible correction within the framework of heterotic string theory.
Recently, we carried out systematic analysis towards realistic models 
within the framework of $SO(32)$ heterotic string theory on the toroidal compactification 
with magnetic fluxes \cite{Abe:2015mua}.
In such a type of model building, we have constructed the models which have  the gauge symmetry 
including  $SU(3) \times SU(2) \times U(1)_Y$ and three generations of quarks and leptons as chiral massless spectra.
Furthermore, in this paper we show that the gauge couplings depend on magnetic fluxes in this type of 
$SO(32)$ heterotic string theory.
That is, gauge couplings can be non-universal 
at the string scale and non-universal parts depend on the K\"ahler moduli, 
although $E_8\times E_8$ heterotic string theory with magnetic fluxes can not lead to 
non-universal gauge couplings between $SU(3)$ and $SU(2)$ appearing in one $E_8$.\footnote{
The low-energy massless spectra were studied within the 10D $E_8$ theory on torus with 
magnetic fluxes from the field-theoretical viewpoint \cite{Choi:2009pv}.} 
Such non-universal corrections can make the gauge coupling prediction consistent with experimental values.
Although such possibilities are proposed in Ref.~\cite{Blumenhagen:2005ga}, 
we study numerically the gauge couplings of $SU(3) \times SU(2) \times U(1)_Y$  
in more realistic models.

This paper is organized as follows.
In section 2, we review 4D low-energy effective field theory 
derived from $SO(32)$ heterotic string theory with magnetic fluxes.
We also review on our model building towards realistic models, which 
have the gauge symmetry including $SU(3) \times SU(2) \times U(1)_Y$ 
and three chiral generations of quarks and leptons as 
well as vector-like matter fields in massless spectra.
In section 3, we study the gauge couplings and show 
how non-universal gauge couplings appear in our models.
In section 4, we study numerically the gauge couplings 
in explicit models.
Section 5 is devoted to conclusions and discussions.

\section{$SO(32)$ heterotic string theory on tori with $U(1)$ magnetic fluxes} 
\label{sec:2}

In this section, we give a review on the low-energy effective field theory 
derived from $SO(32)$ heterotic string theory on factorizable tori with $U(1)$ magnetic fluxes. 
We also show the consistency conditions for $U(1)$ magnetic fluxes 
which give the constraints for the heterotic string models. 
(See for details of model construction e.g., Ref.~\cite{Abe:2015mua}.)

\subsection{Low-energy effective action of $SO(32)$ heterotic string theory} 
\label{subsec:notation}
First of all, we show the bosonic part of $10$D effective supergravity action 
derived from the $SO(32)$ heterotic string theory on a general 
complex manifold $M$ with multiple $U(1)$ magnetic fluxes. 
By calculating the relevant scattering  amplitudes on the worldsheet up to of order 
${\cal O}(\alpha^{\prime})$, we obtain the string-frame bosonic action in the notation of 
Refs.~\cite{Blumenhagen:2005ga,Polchinsky,Weigand:2006yj},
\begin{align}
S_{\rm bos}&=\frac{1}{2\kappa_{10}^2}\int_{M^{(10)}} 
e^{-2\phi_{10}} \left[ R+4d\phi_{10} \wedge  \ast 
d\phi_{10} -\frac{1}{2}H\wedge \ast H \right] 
\nonumber\\
&-\frac{1}{2g_{10}^2} \int_{M^{(10)}} e^{-2\phi_{10}} 
{\rm tr}(F\wedge \ast F),
\label{eq:heterob}
\end{align}
where the gauge and gravitational couplings are 
set by $g_{10}^2 =2(2\pi)^7 (\alpha^{'})^3$ and $2\kappa_{10}^2 =(2\pi)^7 (\alpha^{'})^4$, 
respectively. 
The string coupling is determined by the vacuum expectation 
value of the ten-dimensional dilaton~$\phi_{10}$, that is, 
$g_s=e^{\langle \phi_{10}\rangle}$. 
The field-strength of $SO(32)$ gauge group $F$ has the index of 
vector-representation, which can be normalized as ${\rm tr}_v(T^aT^b)=2\delta^{ab}$. 
In addition, the heterotic three-form field strength $H$ is defined by 
\begin{align}
H&=dB^{(2)} -\frac{\alpha^{'}}{4}(w_{\rm YM} -w_{L}),
\end{align}
where the part of $\alpha^\prime$ corrections are characterized by the gauge and 
gravitational Chern-Simons three-forms, $w_{\rm YM}$ and $w_{L}$, respectively. 

By the ten-dimensional Hodge duality, 
the Kalb-Ramond two-form $B^{(2)}$ and its dual six-form $B^{(6)}$ 
are related as,
\begin{align}
\ast dB^{(2)} =e^{2\phi_{10}} dB^{(6)}.
\label{eq:Bhod}
\end{align}
Then, from the $10$D bosonic action~(\ref{eq:heterob}), we can extract the kinetic term of Kalb-Ramond field,  
\begin{align}
S_{\rm kin} +S_{\rm WZ}&= -\frac{1}{4\kappa_{10}^2}\int_{M^{(10)}} 
e^{2\phi_{10}} dB^{(6)} \wedge \ast dB^{(6)} \nonumber\\ 
&+\frac{\alpha~{'}}{8\kappa_{10}^2}\int_{M^{(10)}}
B^{(6)} \wedge \left( {\rm tr}F^2 -{\rm tr}R^2 -4(2\pi)^2 \sum_a N_a \delta(\Gamma_a) \right),
\label{eq:kinwzd}
\end{align}
where we have added the Wess-Zumino terms induced by the magnetic sources for the Kalb-Ramond field $B^{(6)}$, 
i.e., stacks of $N_a$ five-branes with their tension being $T_5=((2\pi)^5 (\alpha^{'})^3)^{-1}$. 
Note that these heterotic five-branes wrap the holomorphic two-cycles $\Gamma_a$ and their Poinc\'are dual 
four-forms are denoted by $\delta (\Gamma_a)$. 

When we study $SO(32)$ heterotic string theory on three 2-tori, $M=(T^2)_1\times (T^2)_2\times (T^2)_3$, 
the tadpole cancellation is obtained by solving the equation of motion for the Kalb-Ramond field,
\begin{align}
\int_{(T^2)_i\times (T^2)_j}
\left({\rm tr}\bar{F}^2 -4(2\pi)^2 \sum_a N_a 
\delta(\Gamma_a) \right)
=0,
\label{eq:tad2}
\end{align}
where $\bar{F}$ represents for the internal $U(1)$ gauge field strengths. 
These conditions should be satisfied on any four-cycles $(T^2)_i\times (T^2)_j$ 
with $i\neq j$, $i,j=1,2,3$. 
However, if the nonvanishing fluxes are not canceled by themselves, 
five-branes would contribute to the tadpole cancellation~\cite{Witten:1995gx}.

\subsection{Axionic coupling through the Green-Schwarz term}
\label{subsec:GGS}

In addition to the effective action (\ref{eq:heterob}), 
the loop effects induce the Green-Schwarz term 
at the string frame~\cite{Green:1984bx,Ibanez:1986xy}, 
\begin{align}
S_{\rm GS}=\frac{1}{24(2\pi)^5\alpha'}\int B^{(2)}\wedge X_8,
\label{eq:GS}
\end{align}
whose normalization factor is determined by its S-dual type 
I theory as shown in \cite{Blumenhagen:2006ux} 
and the anomaly eight-form $X_8$ reads,
\begin{align}
X_8=\frac{1}{24}{\rm Tr}F^4-\frac{1}{7200}({\rm Tr}F^2)^2-
\frac{1}{240}({\rm Tr}F^2)({\rm tr}R^2) +\frac{1}{8}{\rm tr}R^4 
+\frac{1}{32}({\rm tr}R^2)^2,
\end{align}
where ``Tr" stands for the trace in the adjoint representations of the 
$SO(32)$ gauge group.  

As pointed out in Refs.~\cite{Witten:1984dg,Blumenhagen:2005ga}, 
the gauge and gravitational anomalies for the 
(non-)Abelian gauge groups are canceled 
by the above Green-Schwarz term~(\ref{eq:GS}) and the tadpole 
condition~(\ref{eq:tad2}). 
It is remarkable that 
even if the Abelian gauge symmetries are anomaly-free, 
the Abelian gauge bosons may become massive due to the 
Green-Schwarz coupling given by Eq.~(\ref{eq:GS}). 
Therefore, in order to ensure that the hypercharge 
gauge boson is massless, they should 
not couple to the axions which are hodge dual to the 
Kalb-Ramond fields. 

Let us study the couplings between the hypercharge $U(1)_Y$ gauge boson 
and the axions, explicitly. 
First of all, we decompose the $SO(32)$ 
gauge group into the standard-like model gauge group,
\begin{align}
SO(32) \rightarrow 
SU(3)_C \otimes SU(2)_L \otimes_{a=1}^{13} U(1)_a,
\end{align}
which can be realized by inserting all the multiple $U(1)$ 
constant magnetic fluxes. 

Within the $16$ Cartan elements, $H_i$ ($i=1,\cdots,16$) in $SO(32)$ 
gauge group, 
the Cartan elements of $SU(3)$ are chosen along 
$H_1 -H_2$, $H_1 +H_2 -2H_3$ and that of $SU(2)$ is taken as 
$H_5 - H_6$, whereas the other Cartan directions of $SO(32)$ are 
defined as, 
\begin{align}
&U(1)_1:\,\,\frac{1}{\sqrt{2}}(0,0,0,0,1,1;0,0,\cdots, 0), 
\nonumber\\ 
&U(1)_2:\,\,\frac{1}{2}(1,1,1,1,0,0;0,0,\cdots, 0), 
\nonumber\\
&U(1)_3:\,\,\frac{1}{\sqrt{12}}((1,1,1,-3,0,0;0,0,\cdots, 0),
\nonumber\\
&U(1)_4:\,\,(0,0,0,0,0,0;1,0,\cdots, 0),
\nonumber\\
&U(1)_5:\,\,(0,0,0,0,0,0;0,1,\cdots, 0),
\nonumber\\
& \qquad \vdots 
\nonumber\\
&U(1)_{13}:\,\,(0,0,0,0,0,0;0,0,\cdots, 1),
\label{eq:cartan}
\end{align}
in the basis $H_i$.

Furthermore, when these $U(1)$ fluxes are inserted along the Cartan 
direction of $SO(32)$, the 
field strengths of $U(1)$'s are also decomposed into the four-and extra-dimensional parts 
$f$, $\bar{f}$, respectively. 
Then we can dimensionally reduce the one-loop 
Green-Schwarz term (\ref{eq:GS}) to 
\begin{align}
S_{\rm GS} = \,&\frac{1}{(2\pi )^3 l_s^2}
\int_{M^{(10)}} B^{(2)} \wedge \frac{1}{144}({\rm Tr}F\bar{f}^3)
\label{eq:GSA}\\
&-\frac{1}{(2\pi )^3 l_s^2}
\int_{M^{(10)}} B^{(2)} \wedge \frac{1}{2880}({\rm Tr}F\bar{f}) \wedge 
\left( \frac{1}{15}{\rm Tr}\bar{f}^2 +{\rm tr}{\bar R}^2\right) 
\label{eq:GSB}\\
&+\frac{1}{(2\pi )^3 l_s^2}
\int_{M^{(10)}} B^{(2)} \wedge 
\Bigl[\frac{1}{96}({\rm Tr}F^2\bar{f}^2) 
-\frac{1}{43200}({\rm Tr}F\bar{f})^2\Bigl] 
\label{eq:GSC}\\
&-\frac{1}{(2\pi )^3 l_s^2}
\int_{M^{(10)}} B^{(2)} \wedge 
\frac{1}{5760}({\rm Tr}F^2) \wedge 
\left( \frac{1}{15}{\rm Tr}\bar{f}^2 +{\rm tr}{\bar R}^2\right) 
\label{eq:GSD}\\
&+\frac{1}{(2\pi )^3 l_s^2}
\int_{M^{(10)}} B^{(2)} \wedge 
\frac{1}{384}({\rm tr}R^2) \wedge 
\left( {\rm tr}\bar{R}^2 -\frac{1}{15}{\rm Tr}{\bar f}^2\right), 
\label{eq:GSE}
\end{align}
where $l_s=2\pi \sqrt{\alpha^{\prime}}$ and $F$ denote the 
field strengths of $SU(3)_C$, $SU(2)_L$ and $U(1)_a$, $(a=1,\cdots, 13)$. 

From here, we write the Kalb-Ramond field $B^{(2)}$ 
and internal $U(1)_a$ field strengths $\bar{f}_a$, $(a=1,\cdots, 13)$ 
in the basis of K\"ahler forms $w_i$ on tori $(T^2)_i$ 
\begin{align}
&B^{(2)} =b_S^{(2)} +l_s^2\sum_{i=1}^{3} b_i^{(0)} w_i, 
\nonumber\\
&\bar{f}_a=2\pi d_a \sum_{i=1}^{3} m_a^{(i)} w_i,
\label{eq:expansion}
\end{align}
where $d_a$ are normalization factors appeared in the basis $H_i$~(\ref{eq:cartan}) and $m_a^{(i)}$ are the $U(1)_a$ fluxes constrained by the 
Dirac quantization condition. 
From the Eqs.~(\ref{eq:GSA}) and (\ref{eq:GSB}), we can 
extract the Stueckelberg couplings between the 
$U(1)$ gauge fields and the universal axion 
$b_S^{(0)}$ which is the hodge dual of the tensor field $b_S^{(2)}$ ,
\begin{align}
\frac{1}{3(2\pi )^3 l_s^2} 
\int b_S^{(2)} \wedge &\Bigl[ 
{\rm tr}T_1^4{\bar f}_1^3f_1 +
\left({\rm tr}T_2^4{\bar f}_2^3 +
3({\rm tr}T_2^2T_3^2){\bar f}_2{\bar f}_3^2 +
({\rm tr}T_2T_3^3){\bar f}_3^3 
\right) f_2 
\nonumber\\
&+\left(
{\rm tr}T_3^4{\bar f}_3^3 +
3({\rm tr}T_2T_3^3){\bar f}_2{\bar f}_3^2 +
3({\rm tr}T_2^2T_3^2){\bar f}_2^2{\bar f}_3
\right) f_3 
+\sum_{c=4}^{13}{\rm tr}T_c^4{\bar f}_c^3f_c\Bigl],
\end{align}
which implies that one of the multiple $U(1)$ gauge fields 
absorbs the universal axion and become massive. 
As shown in Sec.~\ref{sec:g-couplings}, 
the hypercharge $U(1)_Y$ is identified with the linear 
combinations of multiple $U(1)$'s, i.e., 
$U(1)_Y=\frac{1}{6}(U(1)_3 +3\sum_{c}U(1)_c)$, 
where the summation over $c$ depends on the 
concrete models.  
In such cases, the $U(1)_Y$ gauge field becomes massless under 
\begin{align}
&6{\rm tr}(T_3^4) m_3^{(1)}m_3^{(2)}m_3^{(3)} 
+3{\rm tr}(T_2T_3^3) d_{ijk}m_2^{(i)}m_3^{(j)}m_3^{(k)}
+3{\rm tr}(T_2^2T_3^2) d_{ijk}m_2^{(i)}m_2^{(j)}m_3^{(k)}
\nonumber\\
&+18\sum_{c} {\rm tr}(T_c^4)m_c^{(1)}m_c^{(2)}m_c^{(3)}=0,
\label{eq:massless1}
\end{align} 
where $d_{ijk}$ denotes the intersection number and there appear  
the non-vanishing intersection numbers of 2-tori, $d_{ijk}=1$ 
($i\neq j\neq k$).

In addition to the universal axion, other axions also appear from the associated 
internal two-cycles, which are known as the K\"ahler axions. 
When the dual field $B^{(6)}$ is expanded as 
\begin{align}
B^{(6)} =l_s^6b_S^{(0)}{\rm vol}(M) 
+l_s^4\sum_{k=1}^{3} b_k^{(2)} \hat{w}_k, 
\end{align}
where $\hat{w}_k$ are the Hodge dual four-forms of the 
K\"ahler forms, 
we can extract the axionic couplings between K\"ahler axions and 
the $U(1)$ gauge bosons through Eq.~(\ref{eq:kinwzd}), 
\begin{align}
&\frac{1}{l_s^2} 
\int b_{i}^{(2)} \wedge \sum_{a=1}^{13}f_a m_{a}^{(i)},
\end{align}
which lead to the following $U(1)_Y$ massless condition, 
\begin{align}
&m_3^{(i)} +3\sum_{c=4}^{13} m_c^{(i)}=0,
\label{eq:massless2}
\end{align} 
with $i=1,2,3$.

\subsection{Model building and constraints}
\label{subsec:4d}
Here we review our approach to construct realistic models.
(See the detail for Ref.~\cite{Abe:2015mua}.)
So far, we introduce the magnetic fluxes $m^{(i)}_a$ along all $U(1)_a$ for $a=1,\cdots,13$.
In our model, such magnetic fluxes break $SO(32)$ into 
$SU(3)_C \times SU(2)_L \times \prod_{a=1}^{13} U(1)_a$ 
and a certain linear combination of $\prod_{a=1}^{13} U(1)_a$ corresponds to $U(1)_Y$.
However, the degenerate magnetic fluxes lead to the enhancement of gauge symmetry e.g., 
$SU(4) \times SU(2) \times SU(2)$ in the visible sector. 
In such a case, we introduce Wilson lines to break this remaining large gauge group 
into $SU(3)_C \times SU(2)_L \times U(1)_Y$.

In addition to gauge symmetry breaking, non-vanishing magnetic fluxes 
can realize the $4$D chiral theory, where the number of zero modes is determined by their 
$U(1)$ charges and magnetic fluxes. 
The relevant matter contents in the SM reside in the adjoint and vector representations of $SO(12)$ 
in $SO(32)$ and their generation numbers are given by
\begin{align}
\begin{array}{ll}
m_{Q_1} =\prod_{i=1}^3 m_{Q_1}^i 
=\prod_{i=1}^3 (m_{1}^i+m_2^i+m_3^i), &
m_{Q_2} =\prod_{i=1}^3 m_{Q_2}^i 
=\prod_{i=1}^3 (-m_{1}^i+m_2^i+m_3^i),\\
m_{L_1} =\prod_{i=1}^3 m_{L_1}^i 
=\prod_{i=1}^3 (m_{1}^i+m_2^i-3m_3^i),&
m_{L_2} =\prod_{i=1}^3 m_{L_2}^i 
=\prod_{i=1}^3 (-m_{1}^i+m_2^i-3m_3^i),\\
m_{u_{R_1}^c} =\prod_{i=1}^3 m_{u_{R_1}^c}^i 
=\prod_{i=1}^3 (-4m_3^i),&
m_{n_{1}} =\prod_{i=1}^3 m_{n_{1}}^i 
=\prod_{i=1}^3 (2m_1^i),\\
m_{d_{R_1}^c} =\prod_{i=1}^3 m_{d_{R_1}^c}^i 
=\prod_{i=1}^3 (2m_2^i+2m_3^i),&
m_{d_{R_2}^c} =\prod_{i=1}^3 m_{d_{R_2}^c}^i 
=\prod_{i=1}^3 (-2m_2^i+2m_3^i),
\\
m_{L_3^a} =\prod_{i=1}^3 m_{L_3^a}^i 
=\prod_{i=1}^3 (m_{1}^i-m_a^i), & 
m_{L_4^a} =\prod_{i=1}^3 m_{L_4^a}^i 
=\prod_{i=1}^3 (-m_{1}^i-m_a^i),\\
m_{u_{R_2}^{c\,a}} =\prod_{i=1}^3 m_{u_{R_2}^{c\,a}}^i 
=\prod_{i=1}^3 (-m_2^i-m_3^i-m_a^i), & 
m_{d_{R_3}^{c\,a}} =\prod_{i=1}^3 m_{d_{R_3}^{c\,a}}^i 
=\prod_{i=1}^3 (-m_2^i-m_3^i+m_a^i),\\
m_{e_{R_1}^{c\,a}} =\prod_{i=1}^3 m_{e_{R_1}^{c\,a}}^i 
=\prod_{i=1}^3 (-m_2^i+3m_3^i+m_a^i), & 
m_{n_2^{a}} =\prod_{i=1}^3 m_{n_2^{a}}^i 
=\prod_{i=1}^3 (-m_2^i+3m_3^i-m_a^i),
\label{eq:gen6612}
\end{array}
\end{align}
where $Q_{1,2}$ are the left-handed quarks, $L_{1,2}, L_{3,4}^a$ are the 
charged leptons and/or Higgs, $u_{R_1}^c, u_{R_2}^{c\,a}$ are the charge 
conjugate of right-handed up type quarks, $d_{R_{1,2}}^c, d_{R_3}^{c\,a}$ 
are the charge conjugate of right-handed down type quarks, $e_{R_1}^{c\,a}$ 
are the charge conjugate of right-handed leptons, $n_1, n_2^a$ are the 
singlets in the standard model gauge groups.

It is remarkable that there are constraints for these $U(1)$ magnetic fluxes. 
First one is that the $U(1)_Y$ massless conditions~(\ref{eq:massless1}) 
and~(\ref{eq:massless2}) by taking account of the axionic couplings with 
$U(1)_Y$ gauge boson. 
Furthermore, there are the tadpole conditions given by Eq.~(\ref{eq:tad2}).
When the heterotic five-branes are absent in our system, the Eq.~(\ref{eq:tad2}) 
is rewritten as 
\begin{align}
&\sum_{a=1}^{13}m_a^{(i)}m_a^{(j)} =0,
\,\,i\neq j,\,\,\,(i,j=1,2,3),
\label{eq:no-five}
\end{align}
which are required from the consistencies of heterotic string theory. 

Without the existence of the heterotic five-branes, it is known that $U(1)$ 
magnetic fluxes satisfy the so-called K-theory constraints, e.g., Ref.~\cite{Blumenhagen:2005ga},
\begin{eqnarray}
\sum_{a=1}^{13} m_a^i = 0 \quad ({\rm mod}~2),
\label{eq:Kth}
\end{eqnarray}
for $i=1,2,3$. 
These K-theory constraints are discussed in the S-dual to the $SO(32)$ 
heterotic string theory, that is, type I string theory. (See for instance, Ref.~\cite{Witten:1998cd,Uranga:2000xp}.)

Finally, non-vanishing magnetic fluxes generically induce the non-vanishing Fayet-Illipoulous (FI) 
terms for $U(1)_a$ with $a=1,2,\cdots, 13$. 
Even if such FI terms are not canceled by themselves, they may be able to be canceled by the vacuum 
expectation values (VEVs) of scalar fields in the hidden sector.

\section{Gauge couplings in heterotic string}
\label{sec:g-couplings}
In this section, we show the formula of gauge kinetic functions in our model. 
After dimensionally reducing the $10$D effective action~(\ref{eq:heterob}) as well as 
the one-loop GS term~(\ref{eq:GS}), 
it is found that the gauge kinetic functions of $SU(3)_C$ and $SU(2)_L$ receive the 
different one-loop threshold corrections depending on the abelian fluxes, while that of 
$U(1)_Y$ do not receive such corrections due to the vanishing axionic couplings with $U(1)_Y$ 
gauge boson.

\subsection{Gauge couplings at tree-level}

\begin{figure}[h]
\centering \leavevmode
\includegraphics[width=0.6\linewidth]{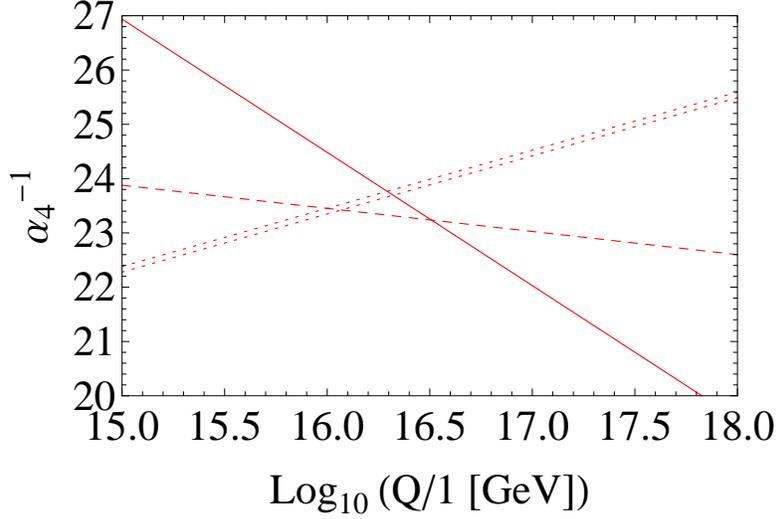}
\caption{RG flow of the gauge couplings in the MSSM 
at the two-loop level. These lines show the gauge coupling 
of $U(1)_Y$ (thick line), $SU(2)_L$ (dashed line) and $SU(3)_C$ 
(dotted line), respectively. Here, we include the error bar associated with 
the QCD coupling $\alpha_{SU(3)_C}^{-1}$~\cite{Agashe:2014kda}.}
\label{fig:rgflow}
\end{figure}

\begin{figure}[h]
\centering \leavevmode
\includegraphics[width=0.6\linewidth]{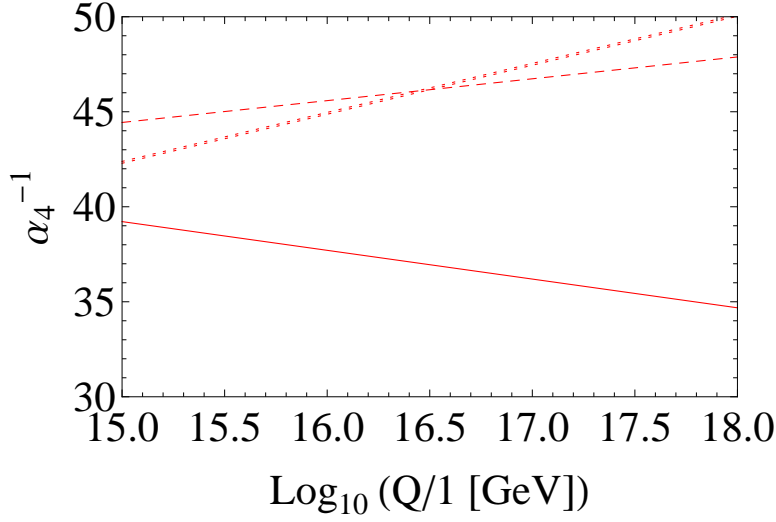}
\caption{RG flow of the gauge couplings in the SM at 
the two-loop level. These lines show the gauge coupling 
of $U(1)_Y$ (thick line), $SU(2)_L$ (dashed line) and $SU(3)_C$ 
(dotted line), respectively.}
\label{fig:rgflowsm}
\end{figure}

After compactifying on a $6$D internal manifold M with volume ${\rm Vol (M)}$, the 
$4$D reduced Planck scale $M_{\rm Pl}$ and the gauge coupling constant $g_4$ can be extracted as 
\begin{align}
&M_{\rm Pl}^2 =
\frac{g_s^{-2}{\rm Vol (M)}}{2\kappa_{10}^2},
\nonumber\\
&g_4^{-2}=g_s^{-2}{\rm Vol (M)}g_{10}^{-2},
\label{eq:}
\end{align}
which lead to the following relation between 
the string scale $M_s=1/l_s$ with $l_s=2\pi\sqrt{\alpha^\prime}$ 
being the typical string length, and 
the Planck-scale,
\begin{align}
&M_s^2=\frac{M_{\rm Pl}^2}{4\pi \alpha_4^{-1}},
\label{eq:stscale}
\end{align}
where $\alpha_4^{-1}=4\pi g_4^{-2}$. 
Since the four-dimensional gauge coupling is 
determined by the VEV
of the dilaton at the tree-level, $\langle{\rm Re}S \rangle=g_4^{-2}$, where 
\begin{align}
S=\frac{1}{2\pi}\left( 
\frac{e^{-2\phi_{10}} {\rm Vol (M)}}{l_s^6} 
+ib_S^{(0)}\right),
\end{align}
with $b_S^{(0)}$ being the universal axion, 
the string scale is roughly estimated as,
\begin{align}
M_s \simeq 1.4 \times 10^{17}\,{\rm GeV},
\end{align}
from Eq.~(\ref{eq:stscale}) by employing $M_{\rm Pl}=2.435\times 10^{18}\,{\rm GeV}$ and 
the four-dimensional gauge coupling constant, 
$\alpha_4^{-1}\simeq 25$, implied by the 
renormalization group (RG) equations of the MSSM. 
As mentioned in the introduction, the gauge 
couplings of SM gauge groups are 
different from each other at the string scale 
as illustrated in 
Fig.~\ref{fig:rgflow} under the assumption that 
the SUSY is broken at the TeV scale and 
the $U(1)_Y$ gauge coupling 
is normalized as satisfying the so-called 
GUT relation. In Fig.~\ref{fig:rgflow}, we involve two-loop effects to the RG equations. 
On the other hand, when the SUSY is broken at 
the string scale, the behavior of gauge couplings 
from the electroweak scale to the string scale 
is obeyed by the renormalization group equations 
of the standard model. 
As seen in Fig.~\ref{fig:rgflowsm}, the gaps between 
the gauge couplings of non-abelian gauge groups 
are better than that of MSSM. 
However, it requires the slight corrections 
to be coincide with the unified one at the string 
scale. 
In this case, the string scale is estimated as $M_s\simeq 1.0 \times 10^{17}$ GeV 
by use of  $\alpha_4^{-1}\simeq 45$. 
We also employ the experimental values such as the gauge coupling of $SU(3)$, 
$\alpha_{SU(3)_C}^{-1}\simeq 0.1184$, the Weinberg angle 
${\rm sin}^2\theta\simeq 0.231$ and the fine-structure constant $\alpha\simeq 1/128$ 
at the electroweak scale~\cite{Agashe:2014kda} in Figs.~\ref{fig:rgflow} and~\ref{fig:rgflowsm}. 

\subsection{The one-loop threshold corrections}
As shown in the previous section, the dilaton $S$ gives 
the universal gauge kinetic function.
However, the Green-Schwarz term $(\ref{eq:GS})$, in particular 
(\ref{eq:GSC}) and (\ref{eq:GSD}), can lead to non-universal gauge kinetic functions~\cite{Blumenhagen:2005ga}.
Indeed, 
we obtain the non-universal axionic couplings,
\begin{align}
S_{\rm GS} \supset 
&\frac{1}{(2\pi )^3}
\sum_{k=1}^3
\int_{M^{1,3}}b_k^{(0)} {\rm tr}_{SU(3)}F^2 
\int_{T^2\times T^2\times T^2} w_k \wedge 
\left( \frac{1}{8}{\rm tr}(T_2^2)\bar{f}_2^2 \right)
\nonumber\\
&+\frac{1}{(2\pi )^3}
\sum_{k=1}^3
\int_{M^{1,3}}b_k^{(0)} {\rm tr}_{SU(2)}F^2 
\int_{T^2\times T^2\times T^2} w_k \wedge 
\left( \frac{1}{4}{\rm tr}(T_1^2)\bar{f}_1^2 \right),
\end{align}
which lead to the non-universal gauge kinetic 
functions. This is because we insert the 
different $U(1)$ fluxes between $U(1)_1$ and 
$U(1)_2$ Cartan directions. Such structures 
are typical in $SO(32)$ heterotic string theory which 
is expected as the S-dual of type I string theory with 
several D-branes. 
However, the non-universal gauge kinetic 
functions for $SU(3)_C$ and $SU(2)_L$ cannot 
be seen in $E_8 \times E_8$ heterotic string theory 
due to the trace identities, 
${\rm Tr}(F^4)=\frac{1}{100}({\rm Tr}(F^2))^2$, 
where $F$ denotes the gauge field strength of 
$E_8$. From this trace identities, Eq.~(\ref{eq:GSC}) is regarded as the type of 
Eq.~(\ref{eq:GSD}). Therefore, the gauge kinetic 
functions of $SU(3)_C$ and $SU(2)_L$ are equal 
to each other. It might be preferred in the 
non-supersymmetric theory such as the standard 
model from the Fig.~\ref{fig:rgflowsm}, although some 
other threshold corrections are required to 
unify the gauge couplings. 

When we define the K\"ahler moduli as 
\begin{align}
T_k=t_k+ib_k^{(0)}, 
\end{align}
where $t_k$ corresponds to the volume of $(T^2)_k$, 
the gauge kinetic functions of the $SU(3)_C$ and 
$SU(2)_L$ become 
\begin{align}
&f_{SU(3)_C}=S+\beta_3^k T^k,
\nonumber\\
&f_{SU(2)_L}=S+\beta_2^k T^k,
\label{eq:gauge-kinetic-f23}
\end{align}
where 
\begin{align}
\beta_3^k =\frac{(d_2)^2}{8\pi}d_{ijk}m_2^im_2^j,\,\,\,
\beta_2^k =\frac{(d_1)^2}{4\pi}d_{ijk}m_1^im_1^j,
\end{align}
with $d_1=\sqrt{2}$ and $d_2=2$. 
Note that threshold corrections depend on the magnetic fluxes.

On the other hand, since $U(1)_Y$ is defined as the 
linear combinations of multiple $U(1)$'s,
\begin{align}
U(1)_Y=\frac{1}{6}\left( U(1)_3+3\sum_{c=4}^N U(1)_c\right),
\label{eq:UY}
\end{align}
the normalization of $U(1)_Y$ is then determined by 
\begin{align} 
&\frac{1}{\alpha_{U(1)_Y}}= \frac{1}{6\alpha_{U(1)_3}} 
+\sum_{c=4}^N\frac{1}{2\alpha_{U(1)_c}}
= \left(\frac{1}{6}+\frac{N-3}{2}\right) 4\pi {\rm Re}\,
\langle S\rangle,
\end{align}
in which the threshold corrections do not appear due to 
the vanishing axionic couplings with $U(1)_Y$ gauge boson, 
and the gauge kinetic function of $U(1)_Y$ is 
extracted as 
\begin{align} 
f_{U(1)_Y}=\left(\frac{1}{6}+\frac{N-3}{2}\right) S.
\label{eq:fU1Y}
\end{align}

\section{Numerical studies in explicit models}

In this section, we show the models satisfying the several consistency conditions 
in section~\ref{subsec:4d}, where the chiral massless spectra in the visible sector 
are just three generations of quarks and leptons without chiral exotics 
and at the same time, 
the experimental values of gauge couplings are realized at the string scale. 
Although there are extra vector-like visible matter fields and hidden chiral and vector-like matter fields, 
we assume that these modes become massive around $M_s$ such that 
the massless spectra of the SM or the 
MSSM are realized around $M_s$. 

From now on, we consider two concrete scenarios.
In one scenario, supersymmetry is broken at $M_s$ and 
below $M_s$ the massless spectrum in the visible sector is just one of 
the SM.
In other scenario, supersymmetry remains at $M_s$ and breaks 
around $1$ TeV, and below $M_s$ the massless spectrum in the visible sector is just one of 
the MSSM. 
We assume that the non-vanishing D-terms for extra $U(1)$'s generated by the generic magnetic 
flux background are canceled by VEVs of hidden scalar fields 
and ${\cal N}=1$ supersymmetry remains when we take low-energy 
SUSY breaking scenario. 
Since the size of SUSY breaking scale depends on the moduli stabilization scenario, 
we leave the details of them for future work.

First of all, the $U(1)_Y$ gauge coupling $g^2_{U(1)_Y}$ at $M_s$ is determined 
only by $\langle S \rangle$ and $N$ through Eq.~(\ref{eq:fU1Y}),
\begin{align} 
\frac{1}{g^2_{U(1)_Y}(M_s)}= A(N) \langle S \rangle,
\end{align}
where 
\begin{eqnarray}
A(N) = \left(\frac{1}{6}+\frac{N-3}{2}\right).
\end{eqnarray}
From the experimental values of $U(1)_Y$ gauge coupling, 
$g^{-2}_{U(1)_Y}(M_s) = 4.80 $ for the SM and $2.44$ for the MSSM, 
the dilaton VEV $\langle S \rangle$ is determined by $N$ as shown in 
Table \ref{table:dilaton} for $N=5,7,9$.
In what follows, we discuss the explicit models with $N=5,7,9$.

\begin{table}[ht]
 \begin{center}
  \begin{tabular}{|c|c|c|c|} \hline
   $N$ & 5 & 7 & 9 \\ \hline
   SM & 4.11 & 2.22 & 1.52 \\ \hline
   MSSM & 2.47 & 1.33 & 0.91 \\ \hline
  \end{tabular}
 \end{center}
 \caption{The VEV of dilaton $\langle S\rangle$ for the SM and MSSM.}
 \label{table:dilaton}
\end{table}

Next, by solving 
 Eq.~(\ref{eq:gauge-kinetic-f23}) and two-loop renormalization group equations for 
SM and MSSM, we evaluate the ratio of gauge couplings at $M_s$ as
\begin{equation}
 \begin{split}
  &\frac{\langle{\rm Re}f_1\rangle}{\langle{\rm Re}f_3\rangle} 
   =\frac{A(N)}{1+\beta_3^{k}\langle T_k\rangle /\langle S\rangle}
   =\frac{g_3^2(M_s)}{g_1^2(M_s)}\simeq 0.881, \\
  &\frac{\langle{\rm Re}f_1\rangle}{\langle{\rm Re}f_2\rangle} 
   =\frac{A(N)}{1+\beta_2^{k}\langle T_k\rangle /\langle S\rangle}
   =\frac{g_2^2(M_s)}{g_1^2(M_s)}\simeq 0.944,
 \end{split}
 \label{eq:expMSSM}
\end{equation}
for the MSSM,
\begin{equation}
 \begin{split}
  &\frac{\langle{\rm Re}f_1\rangle}{\langle{\rm Re}f_3\rangle} 
   =\frac{A(N)}{1+\beta_3^{k}\langle T_k\rangle /\langle S\rangle}
   =\frac{g_3^2(M_s)}{g_1^2(M_s)}\simeq 0.763, \\
  &\frac{\langle{\rm Re}f_1\rangle}{\langle{\rm Re}f_2\rangle} 
   =\frac{A(N)}{1+\beta_2^{k}\langle T_k\rangle /\langle S\rangle}
   =\frac{g_2^2(M_s)}{g_1^2(M_s)}\simeq 0.775,
 \end{split}
 \label{eq:expSM}
\end{equation}
for the SM. Here, the experimental values such as the gauge coupling of $SU(3)$, 
$\alpha_{SU(3)_C}^{-1}\simeq 0.1184$, the Weinberg angle 
${\rm sin}^2\theta\simeq 0.231$ and the fine-structure constant $\alpha\simeq 1/128$ 
at the electroweak scale are employed. 
From  Eqs.~(\ref{eq:expMSSM}) and (\ref{eq:expSM}), we estimate the following equalities 
in Tab.~\ref{table:B_i} for $N=5,7,9$, 
\begin{equation}
 \begin{split}
   B_2=2\pi\beta_2^k \langle T_k\rangle =m_1^2m_1^3\langle T_1\rangle 
+m_1^3m_1^1\langle T_2\rangle +m_1^1m_1^2\langle T_3\rangle, \\
   B_3=2\pi\beta_3^k \langle T_k\rangle =m_2^2m_2^3\langle T_1\rangle 
+m_2^3m_2^1\langle T_2\rangle +m_2^1m_2^2\langle T_3\rangle.
 \end{split}
 \label{eq:Bi}
\end{equation}
Tab.~\ref{table:B_i} shows that for $N=5$, the values of $|B_2|$ and $|B_3|$ are much 
smaller than ${\cal O}(1)$. 
In order to realize $\langle T_k\rangle ={\cal O}(1)$, i.e., the small value of string coupling, 
it requires the certain cancellations within the 
VEVs of K\"ahler moduli $\langle T_k\rangle $ and $U(1)_{1,2}$ fluxes in Eq.~(\ref{eq:Bi}). 
A similar behavior on $B_2$ and $m_1^i$ would be required for $N=7$. 
On the other hand, for $N=9$, the large $B_3$ requires the large $U(1)_2$ fluxes 
$m_2^i$ to obtain $\langle T_k\rangle ={\cal O}(1)$.

\begin{table}[h]
 \begin{center}
  \begin{tabular}{|c|r|r|r|r|r|r|r|r|} \hline
    & \multicolumn{2}{c|}{Model $1$ ($N=5$)} & \multicolumn{2}{c|}{Model $2$ ($N=7$)} & \multicolumn{2}{c|}{Model $3$ ($N=9$)} \\ \cline{2-7}
    & \multicolumn{1}{c|}{SM} & \multicolumn{1}{c|}{MSSM} & \multicolumn{1}{c|}{SM} & \multicolumn{1}{c|}{MSSM} & \multicolumn{1}{c|}{SM} & \multicolumn{1}{c|}{MSSM} \\ \hline
   $B_2$ & $-0.148$ & $-0.658$ & 1.921 & 1.783 & 6.017 & 6.996 \\ \hline
   $B_3$ & $-0.252$ & $-1.046$ & 3.992 & 4.496 & 12.35 & 15.98 \\ \hline
  \end{tabular}
 \end{center}
 \caption{The values of $B_2$ and $B_3$ for the SM and MSSM in the case of $N=5,7,9$. 
By increasing the number of $N$ appeared in the definition of $U(1)_Y$~(\ref{eq:UY}), 
the values of $B_i$ become larger in both cases of SM and MSSM.}
 \label{table:B_i}
\end{table}

When we construct an explicit model, 
all $U(1)$ magnetic fluxes as well as $N$ appeared in the definition of $U(1)_Y$ given by Eq.~(\ref{eq:UY}) are fixed 
and the $\beta_2^k$ and $\beta_3^k$ in Eq.~(\ref{eq:Bi}) are also fixed hereafter.
Then, we can examine whether the ${\cal O}(1)$ values of $\langle T_k \rangle$ are consistent with the 
values of $B_2$ and $B_3$ in Table~\ref{table:B_i} or not. 
Although it is expected that there appear one of the unfixed K\"ahler moduli by 
solving the two equations~(\ref{eq:Bi}) under the three K\"ahler moduli $\langle T_k \rangle$, 
in some models there are no solutions for realistic values of $T_k$, i.e., 
$T_k<0$, $T_k\ll 1$, $T_k\gg 1$ and so on. 
In addition to it, when one of $m_1^i$ and $m_2^i$ vanishes, all K\"ahler moduli are completely fixed 
by Eq.~(\ref{eq:Bi}). 

In the following, we show the three examples of magnetic flux configurations denoted by 
models $1$ ($N=5$), $2$ ($N=7$) and $3$ ($N=9$) which are realistic in the sense that 
there are the gauge symmetry including $SU(3)_C \times SU(2)_L \times U(1)_Y$, 
three chiral generations of quarks and leptons without chiral exotics in the visible sector, 
the experimental values of gauge couplings in Eq.~(\ref{eq:Bi}) and 
at the same time, they satisfy the consistency conditions in Sec.~\ref{subsec:4d}. 
The procedure of searching for these models are given as follows. 
First, in the light of $U(1)_Y$ massless conditions~(\ref{eq:massless1}) and~(\ref{eq:massless2}), 
we restrict ourselves to the magnetic flux configurations,
\begin{equation}
	\begin{array}{l}
		m_3^i = 0, \\
		m_{3 + a}^i = - m_{8 + a}^i \; (a= 1,2,3,4,5).
	\end{array}
	\label{eq:magbac}
\end{equation}
Next, we classify the $U(1)_{1,2}$ magnetic fluxes $m_{1,2}^i$ so that the three-generations 
of left-handed quarks $Q_{1,2}$ and single-valued wavefunction for the singlet $n_1$ are achieved, 
\begin{align}
&m_{Q_1}+m_{Q_2}=\prod_{i=1}^3 (m_{1}^i+m_2^i) +\prod_{i=1}^3 (-m_{1}^i+m_2^i)=3,\nonumber\\
&m_{n_1}^i=2m_1^i, 
\end{align}
which constrain the $m_{1,2}^i$ as the integers or half-integers. 
Furthermore, from the K-theory condition~(\ref{eq:Kth}) with the magnetic flux background~(\ref{eq:magbac}), 
the generation of left-handed quarks are determined by $m_{Q_1}=0$ and $m_{Q_2}=3$ 
corresponding to the model ``${\rm Type\,B'}$" in Ref.~\cite{Abe:2015mua} due to the 
even numbers of $m_1^i+m_2^i$ for $i=1,2,3$. 
Then, the magnetic fluxes $m_{1,2}^i$ should become half-integers. 

Thus, from the obtained list for $m_{1,2}^i$, we search for the ${\cal O}(1)$ values of K\"ahler 
moduli $\langle T_i\rangle$, $i=1,2,3$ by solving the Eq.~(\ref{eq:Bi}). 
When $m^3_1m_2^2 - m_1^2m_2^3 \neq 0$, 
the K\"ahler moduli $T_{2,3}$ are represented by 
\begin{align}
&T_2=a_2 T_1+b_2,\nonumber\\
&T_3=a_3 T_1+b_3,
\end{align}
where $a_{2,3}$ and $b_{2,3}$ are the flux-dependent constants given through Eq.~(\ref{eq:Bi}), i.e.
\begin{eqnarray}
& & a_2=\frac{m_1^2m_2^2(m^1_1m_2^3 - m_1^3m_2^1)}{m_1^1m_2^1(m^3_1m_2^2 - m_1^2m_2^3)}, \qquad 
b_2 = \frac{m_2^1m_2^2B_2 - m_1^1m_1^2B_3}{m_1^1m_2^1(m^3_1m_2^2 - m_1^2m_2^3)},  \nonumber \\
& & a_3=\frac{m_1^3m_2^3(m^1_1m_2^3 - m_1^3m_2^1)}{m_1^1m_2^1(m^2_1m_2^3 - m_1^3m_2^2)}, \qquad 
b_3 = \frac{m_2^1m_2^3B_2 - m_1^1m_1^3B_3}{m_1^1m_2^1(m^2_1m_2^3 - m_1^3m_2^2)}.  
\end{eqnarray}

\begin{figure}
\centering \leavevmode
\includegraphics[width=90mm]{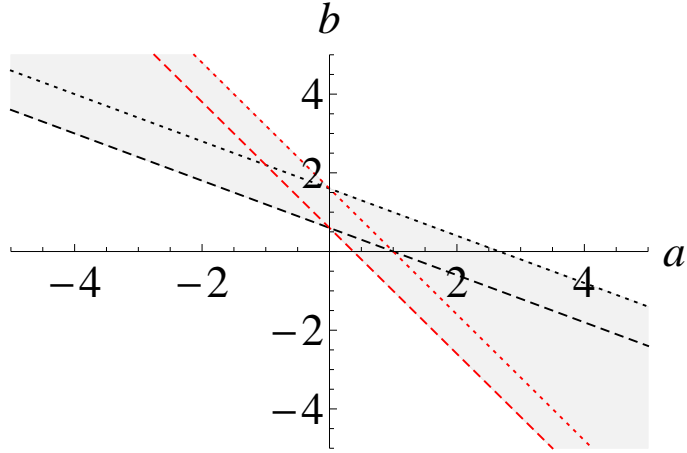}
\caption{The shaded gray region in the parameter spaces ($a,b$) is consistent with ${\cal O}(1)$ values of K\"ahler moduli, 
$T_{\rm min} \leq \langle T_i\rangle \leq T_{\rm max}$, where the lower (upper) bounds are chosen, for instance, $T_{\rm min}=0.6$ ($T_{\rm max}=1.6$). 
The black dashed and dotted lines denote $b=-T_{\rm min}a+T_{\rm min, max}$, 
while the red dashed and dotted lines denote $b=-T_{\rm max}a+T_{\rm min, max}$, 
respectively.}
\label{fig:ab}
\end{figure}

The ${\cal O}(1)$ values of K\"ahler moduli $\langle T_i\rangle$, $i=1,2,3$ 
require the shaded gray parameter region for $a\equiv a_{2,3}$ and $b\equiv b_{2,3}$ as shown in Fig.~\ref{fig:ab}. 
As for the parameter spaces ($a,b$) within the shaded region in Fig.~\ref{fig:ab} corresponding to the 
${\cal O}(1)$ values of K\"ahler moduli, 
we determine the other $U(1)_a$ $a=4,5,\cdots,13$ fluxes so as to achieve the matter contents in 
the standard model in Eq.~(\ref{eq:gen6612}) and 
$U(1)_Y$ massless conditions~(\ref{eq:massless1}) as well as~(\ref{eq:massless2}) and K-theory 
condition~(\ref{eq:Kth}). 

As a result, within the range $2m_1^i \in [-11,11]$ and $0.7\leq \langle T_i\rangle\leq 1.4$, $i=1,2,3$, 
the magnetic flux configurations in model $1$ with and 
without heterotic five-branes are uniquely determined as shown in Tab.~\ref{tab:Model-1}. 
Tabs.~\ref{tab:Model-2} and \ref{tab:Model-3} also show the 
magnetic flux configurations in models $2$ and $3$ with and 
without heterotic five-branes within the range $2m_1^i \in [-11,11]$ and 
$0.8\leq \langle T_i\rangle\leq 1.3$, $i=1,2,3$, respectively. 
In the case without five-branes in Tabs.~\ref{tab:Model-1}, \ref{tab:Model-2} and \ref{tab:Model-3}, 
Eq.~(\ref{eq:no-five}) is satisfied by the $U(1)$ fluxes themselves. 
Moreover, the magnetic fluxes in Tabs.~\ref{tab:Model-1}, \ref{tab:Model-2} and \ref{tab:Model-3} 
predict the same values of gauge couplings at the string scale, because 
only $m_1^i$ and $m_2^i$ appear in the non-universal terms of 
$SU(2)_L$ and $SU(3)_C$ gauge couplings. 

\begin{figure}[h]
\begin{minipage}{0.5\hsize}
\begin{center}
\includegraphics[width=0.8\linewidth]{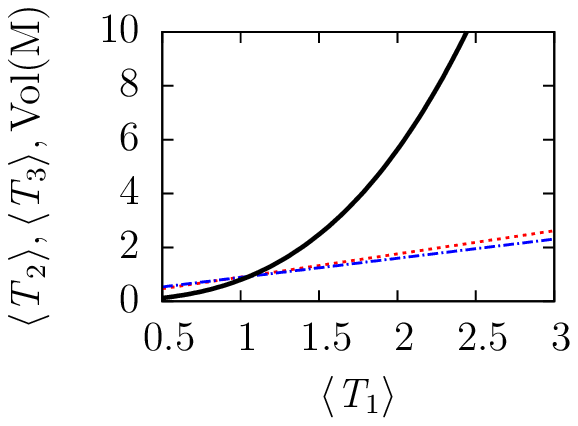}
\end{center}
\end{minipage}
\begin{minipage}{0.5\hsize}
\begin{center}
\includegraphics[width=0.8\linewidth]{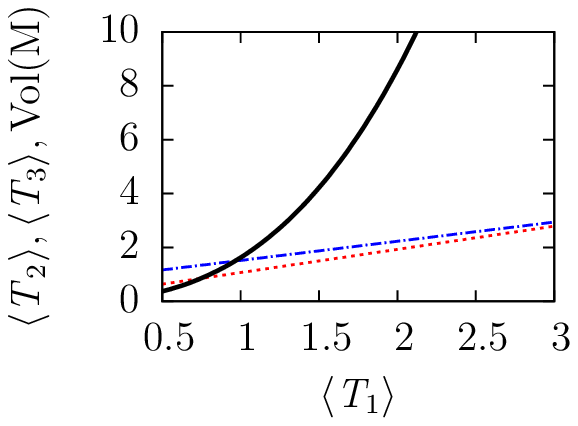}
\end{center}
\end{minipage}
\caption{The VEVs of moduli, $\langle T_2\rangle$ (red dashed curve), 
$\langle T_3\rangle$ (blue dotdashed curve) 
and the volume of three-tori ${\rm Vol (M)}=\langle T_1T_2T_3\rangle$ (black thick curve) as a function 
of $\langle T_1\rangle$ in model $1$. 
The left and right panels show those of SM and MSSM, respectively.}
\label{fig:mod1}
\end{figure}

\begin{figure}[h]
\begin{minipage}{0.5\hsize}
\begin{center}
\includegraphics[width=0.8\linewidth]{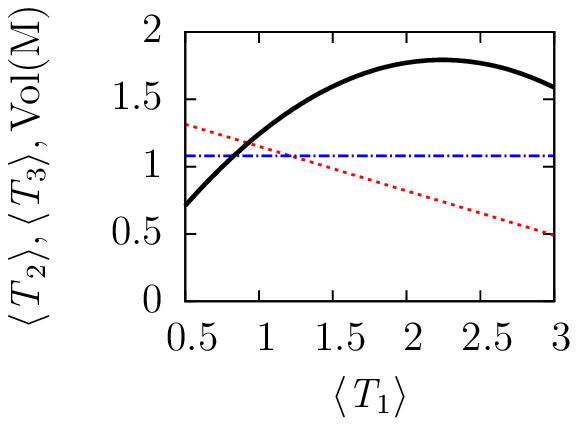}
\end{center}
\end{minipage}
\begin{minipage}{0.5\hsize}
\begin{center}
\includegraphics[width=0.8\linewidth]{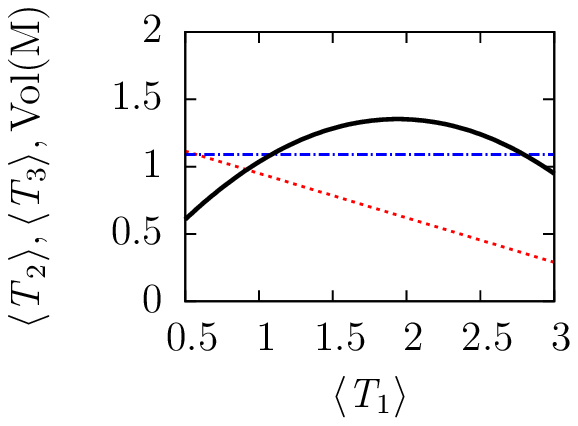}
\end{center}
\end{minipage}
\caption{The VEVs of moduli, $\langle T_2\rangle$ (red dashed curve), 
$\langle T_3\rangle$ (blue dotdashed curve) 
and the volume of three-tori ${\rm Vol (M)}=\langle T_1T_2T_3\rangle$ (black thick curve) as a function 
of $\langle T_1\rangle$ in model $2$. 
The left and right panels show those of SM and MSSM, respectively.}
\label{fig:mod2}
\end{figure}

\begin{figure}[h]
\begin{minipage}{0.5\hsize}
\begin{center}
\includegraphics[width=0.8\linewidth]{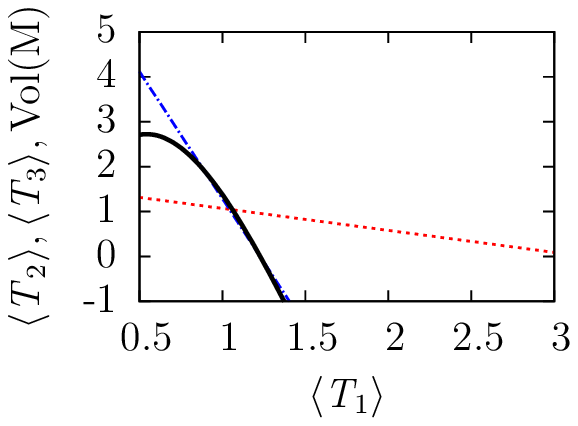}
\end{center}
\end{minipage}
\begin{minipage}{0.5\hsize}
\begin{center}
\includegraphics[width=0.8\linewidth]{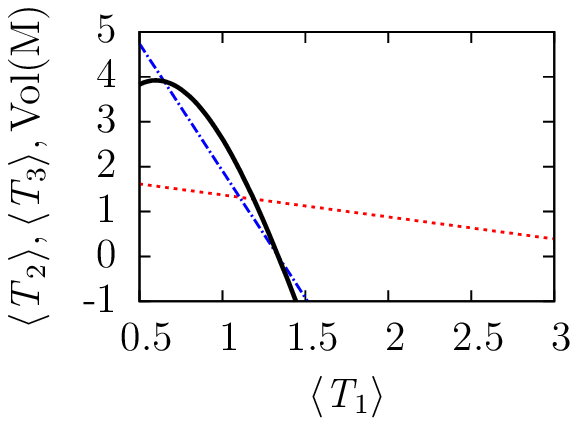}
\end{center}
\end{minipage}
\caption{The VEVs of moduli, $\langle T_2\rangle$ (red dashed curve), 
$\langle T_3\rangle$ (blue dotdashed curve) 
and the volume of three-tori ${\rm Vol (M)}=\langle T_1T_2T_3\rangle$ (black thick curve) as a function 
of $\langle T_1\rangle$ in model $3$. 
The left and right panels show those of SM and MSSM, respectively.}
\label{fig:mod3}
\end{figure}

Finally, the Figs.~\ref{fig:mod1}, \ref{fig:mod2} and \ref{fig:mod3} show the values of K\"ahler moduli and 
volume of three-tori ${\rm Vol (M)}=\langle T_1T_2T_3\rangle$ as a function of $\langle T_1\rangle$ 
in models $1$, $2$ and $3$, respectively. 
The ${\cal O}(1)$ values of K\"ahler moduli in Figs.~\ref{fig:mod1}, \ref{fig:mod2} and \ref{fig:mod3} are 
consistent with the experimental values of gauge couplings in the SM and MSSM. 
Similarly, we can analyze a wider region of 
$T_{\rm min} < \langle T_i\rangle< T_{\rm max} $ with larger $T_{\rm max}$ and 
smaller $T_{\rm min}$.
For example, for $0.3 < \langle T_i\rangle< 3.0 $, there are many models consistent with the 
experimental results of gauge couplings in the SM and MSSM.

\begin{table}[htbp]
	\begin{center}
			\begin{tabular}{|c|c|c|} \hline
			Magnetic fluxes & Without five-branes & With five-branes   \\ \hline
			$(2 m_1^1 , 2 m_1^2 , 2 m_1^3)$ & (1,-3,-1) & (1,-3,-1)  \\
			$(2 m_2^1 , 2 m_2^2 , 2 m_2^3)$ & (7,-1,1) & (7,-1,1) \\
			$(2 m_3^1 , 2 m_3^2 , 2 m_3^3)$ & (0,0,0)  & (0,0,0)\\
			$(2 m_4^1 , 2 m_4^2 , 2 m_4^3)$ & (-5,-1,5) & (-5,-5,1) \\
			$(2 m_5^1 , 2 m_5^2 , 2 m_5^3)$ & (3,1,1) & (1,1,1)\\
			$(2 m_6^1 , 2 m_6^2 , 2 m_6^3)$ & (3,1,1) & (1,1,1) \\
			$(2 m_7^1 , 2 m_7^2 , 2 m_7^3)$ & (3,1,1) & (1,1,1)\\
			$(2 m_8^1 , 2 m_8^2 , 2 m_8^3)$ & (3,1,1) & (1,1,1)\\
			\hline
			\end{tabular}
   \end{center}		
\caption{The magnetic flux configurations in model $1$ with and without five-branes.}	
			\label{tab:Model-1}
\end{table}

\begin{table}[htbp]
	\begin{center}
			\begin{tabular}{|c|c|c|} \hline
			 Magnetic fluxes & Without five-branes & With five-branes   \\ \hline			
			$(2 m_1^1 , 2 m_1^2 , 2 m_1^3)$ & (-3,-1,-1) & (-3,-1,-1) \\
			$(2 m_2^1 , 2 m_2^2 , 2 m_2^3)$ & (-9,-3,1) & (-9,-3,1)  \\
			$(2 m_3^1 , 2 m_3^2 , 2 m_3^3)$ & (0,0,0)   & (0,0,0)  \\
			$(2 m_4^1 , 2 m_4^2 , 2 m_4^3)$ & (11,-1,1) & (9,-1,1) \\
			$(2 m_5^1 , 2 m_5^2 , 2 m_5^3)$ & (11,1,1) & (9,-1,1) \\
			$(2 m_6^1 , 2 m_6^2 , 2 m_6^3)$ & (5,-3,-1)   & (9,-1,1) \\
			$(2 m_7^1 , 2 m_7^2 , 2 m_7^3)$ & (9,-1,1)  & (9,-1,1)\\
			$(2 m_8^1 , 2 m_8^2 , 2 m_8^3)$ & (11,-3,-1)  & (9,-1,1)\\
			\hline
			\end{tabular} 
   \end{center}
   \caption{The magnetic flux configurations in model $2$ with and without five-branes.}
					\label{tab:Model-2}	
\end{table}
		
\begin{table}[htbp]
	\begin{center}
			\begin{tabular}{|c|c|c|} \hline
			Magnetic fluxes & Without five-branes & With five-branes   \\ \hline
			$(2 m_1^1 , 2 m_1^2 , 2 m_1^3)$ & (1,1,11) & (1,1,11)  \\
			$(2 m_2^1 , 2 m_2^2 , 2 m_2^3)$ & (7,-1,9) & (7,-1,9) \\
			$(2 m_3^1 , 2 m_3^2 , 2 m_3^3)$ & (0,0,0)  & (0,0,0)\\
			$(2 m_4^1 , 2 m_4^2 , 2 m_4^3)$ & (-5,-1,-7) & (-5,-1,-7) \\
			$(2 m_5^1 , 2 m_5^2 , 2 m_5^3)$ & (-5,-1,-7) & (-5,-1,-7)\\
			$(2 m_6^1 , 2 m_6^2 , 2 m_6^3)$ & (-5,-1,-7) & (-5,-1,-7) \\
			$(2 m_7^1 , 2 m_7^2 , 2 m_7^3)$ & (7,1,-23) & (7,1,-3)\\
			$(2 m_8^1 , 2 m_8^2 , 2 m_8^3)$ & (9,-1,-9) & (7,1,-3)\\
			\hline
			\end{tabular} 
   \end{center}			
   \caption{The magnetic flux configurations in model $3$ with and without five-branes.}
			\label{tab:Model-3}
\end{table}	


\clearpage
\section{Conclusion}

We have studied on $SO(32)$ heterotic models with $U(1)$ magnetic fluxes, which 
have the gauge symmetry including $SU(3)_C \times SU(2)_L \times U(1)_Y$ and 
three chiral generations of quarks and leptons as well as vector-like 
matter fields. 
In contrast to $E_8\times E_8$ heterotic string theory, 
there is the non-universality among the gauge couplings of standard model at the string scale 
and they depend on magnetic fluxes as well as the VEVs of 
dilaton and K\"ahler moduli. 
Although there are several approaches to realize the gauge couplings 
consistent with their experimental values, they require the large stringy threshold corrections by 
employing the large field values of K\"ahler moduli~\cite{Dixon:1990pc,Antoniadis:1991fh,Derendinger:1991hq} 
which implies the large string coupling at the vacuum. 
In this paper, we have considered the two SUSY breaking scenarios. 
One of them is that the SUSY is broken at the string scale, whereas 
the other model is the TeV SUSY breaking scenario. 
In both scenarios, it was found that certain explicit models can lead to 
the gauge couplings consistent with the experimental values even if 
the values of K\"ahler moduli are of order unity. 
Thus, we have constructed the realistic models from both viewpoints 
of massless spectra and the gauge couplings. 

What is important for a next study would be Yukawa couplings.
The zero-mode profiles of quarks and leptons as well as higgs fields 
are non-trivial because of introducing magnetic fluxes.
That would lead to non-trivial Yukawa matrices.\footnote{
See  for a relevant study on magnetized brane models, e.g. Refs.~\cite{Cremades:2004wa,Abe:2008sx}.}.
Also, in Ref.~\cite{Abe:2015mua}, 
it was shown that the models with $N=9$ have $SU(3)$ flavor symmetry.
Such a flavor symmetry might be useful to realize the realistic values of
fermion masses and mixing angles.
We would study this issue elsewhere.

So far, we have taken the dilaton and K\"ahler moduli VEVs as 
free parameters in order to obtain the gauge couplings 
consistent with the experimental values. 
It is also next issue to study moduli stabilization at proper values of them.

\subsection*{Acknowledgement}
H.~A. was supported in
part by the Grant-in-Aid for Scientific Research No. 25800158 from the
Ministry of Education,
Culture, Sports, Science and Technology (MEXT) in Japan. T.~K. was
supported in part by
the Grant-in-Aid for Scientific Research No. 25400252 from the MEXT in
Japan.
H.~O. was supported in part by a Grant-in-Aid for JSPS Fellows 
No. 26-7296.

\end{document}